\documentclass[twocolumn,preprintnumbers,amsmath,amssymb,nofootinbib,superscriptaddress]{revtex4}

\usepackage{epsfig}
\usepackage{amsfonts}
\usepackage{dcolumn}
\usepackage{dsfont}
\usepackage{epsfig}
\usepackage{subfigure}

\newcommand{\TeV}{\,\mathrm{TeV}}
\newcommand{\GeV}{\,\mathrm{GeV}}

\newcommand{\nn}{\nonumber}

\newcommand{\redh}{{ h}}
\newcommand{\redeta}{{S}}
\newcommand{\vev}[1]{\langle {#1} \rangle}
\newcommand{\pslash}{\!\not\! p}
\def\simlt{\stackrel{<}{{}_\sim}}

\usepackage{amsmath,amssymb}

\newcommand{\be}{\begin{equation}}
\newcommand{\ee}{\end{equation}}
\def\bea{\begin{eqnarray}}
\def\eea{\end{eqnarray}}
\def\Z2{\mathbf{Z}_2}

\begin{document}

\title{Electroweak Baryogenesis in Non-minimal Composite Higgs Models} 

\date{\today}

\author{Jos\'e R. Espinosa}  
\affiliation{ICREA, Instituci\`o Catalana de Recerca i Estudis Avan\c{c}ats,
Barcelona, Spain.}
\affiliation{IFAE, Universitat Aut{\`o}noma de Barcelona,
08193 Bellaterra, Barcelona, Spain.}

\author{Ben Gripaios}  
\affiliation{Cavendish Laboratory, University of Cambridge,
  J. J. Thomson Avenue, Cambridge CB3 0HE, United Kingdom.}

\author{Thomas Konstandin}
\affiliation{CERN PH-TH, Geneva 23, 1211 Switzerland.}

\author{Francesco Riva}  
\affiliation{IFAE, Universitat Aut{\`o}noma de Barcelona,
08193 Bellaterra, Barcelona, Spain.}
\affiliation{Istituto Galileo Galilei, Universit\'a degli studi di Padova, Via Marzolo 
 8, I-35131 Padua, Italy.}

\preprint{}
\begin{abstract}
We address electroweak baryogenesis in the context of composite Higgs
models, pointing out that modifications to the Higgs and top quark sectors can play an important r\^ole in generating the baryon asymmetry. Our main observation is that composite Higgs models that include a  light, gauge singlet scalar in the spectrum [as in the model based on
the symmetry breaking pattern $SO(6) \to SO(5)$], provide all necessary
ingredients for viable baryogenesis. In particular, the singlet leads to a strongly first-order electroweak phase transition and introduces new
sources of $CP$ violation in dimension-five operators involving the
top quark. We discuss the amount of  baryon asymmetry produced and the experimental constraints on the model.
\end{abstract}   
\maketitle

\section{Motivation}

It is well known that the baryon asymmetry observed in the universe
cannot be explained by the Standard Model (SM) of particle
physics. Indeed, neither the $CP$ violation encapsulated in the CKM
matrix nor the departure from thermal equilibrium during the
electroweak phase transition (EWPhT) are large enough, given the LEP lower bound on the mass of the Higgs boson \cite{LEPmhSM}. This calls for physics beyond the Standard Model. In particular, new bosons could enhance the strength of the EWPhT, while new fermions, or new fermionic couplings, could provide additional sources of $CP$ violation. 

Both ingredients are present in supersymmetric extensions of the SM and (perhaps not surprisingly given the added motivations coming from naturalness, gauge coupling unification, and Dark Matter) most studies of electroweak baryogenesis (EWBG) have concentrated on the minimal supersymmetric standard model (MSSM) \cite{MSSMEWBG}. Unfortunately, the LEP bound on the Higgs mass \cite{LEPmhMSSM} has pushed MSSM EWBG into fine-tuned territory
\cite{Carena:2002ss, Konstandin:2005cd, Carena:2008vj,
Cirigliano:2009yd, DeSimone:2011ek}. Furthermore, this and other
bounds on sparticle masses from LEP and LHC \cite{SUSYCMS,SUSYATLAS,Strumia} have set strong constraints on some scenarios of low-energy supersymmetry, putting the
original naturalness motivation under some pressure.  Although
non-minimal supersymmetric extensions of the MSSM can overcome some of
these problems \cite{NotMSSMEWBG,Davies:1996qn,Huber:2000mg,Kang:2004pp,Menon:2004wv,Fok:2008yg,Chung:2010cd,Kanemura:2011fy,BMSSMEWBG}, the time is
perhaps right to examine EWBG in non-supersymmetric models that address the hierarchy problem (and necessarily incorporate new degrees of freedom at the EW scale).

Among the possibilities that have been studied, models in which the
Higgs boson is composite and the SM fermions are partially composite
seem to offer the most plausible alternative to supersymmetry.  In such models \cite{CH,CH:ACP} the Higgs boson arises as a bound state of new,
strongly-interacting dynamics and is naturally light, since it
originates as a pseudo Nambu-Goldstone Boson (PNGB) of a broken global
symmetry, similar to the pions of QCD.  The pattern of symmetry
breaking can be chosen to reduce the tension with the electroweak
precision observables (EWPO) $T$ \cite{CH:ACP} and $Zb\bar b$
\cite{Agashe:2006at}, while the presence of a (composite) Higgs boson
allows a mechanism for suppressing the contributions to
$S$. Moreover, the assumption that the SM fermions are partially
composite \cite{Kaplan:1991dc} results in a suppression of dangerous
contributions to flavor physics observables, provided the theory is
approximately conformal in the ultraviolet \cite{Huber:2000ie,Gherghetta:2000qt, PartComp}.

Such a theory would presumably give rise to a plethora of bosonic and fermionic
resonances at or around the strong coupling scale and this exotic
dynamics could certainly play a r\^ole in generating the observed baryon
asymmetry (ref.~\cite{Konstandin:2011ds}, for example, pointed out that the radion, in the context of the AdS/CFT correspondence, could be relevant).  Indeed it would seem that the connection between
naturalness arguments and EWBG would be especially tight in such a
theory, since the dominant radiative corrections (involving the
mostly-composite Higgs and top quark sectors) shape the potential at
both zero and non-zero temperatures and would be responsible both for
stabilizing the weak scale and generating the required
strongly first-order phase transition. Moreover, new, strong couplings
between the top and Higgs sectors could provide the required source of
$CP$-violation.

Unfortunately, a quantitative demonstration of this link would seem to be out of
the question if the strongly-coupled physics lies close to
the EW scale, given our limited understanding of
strongly-coupled dynamics. However, current data suggest that the
strong coupling scale must be pushed up to a few TeV or so.  Indeed, the suppression of the $S$
parameter mentioned above is achieved in precisely this way and a
similar hierarchy is also needed for sufficient suppression of
contributions to flavor physics observables. Whilst this separation is
bad news for naturalness (it requires a fine-tuning at the
level of $10\%$ or so), it at least offers the hope that we may
be able to study EWBG in such models in an effective theory approach,
in which the details of the strong dynamics at the strong coupling
scale are integrated out. In such a framework, one only needs to
specify the light degrees of freedom, together with the operators of
low dimension that are present in the low-energy effective
Lagrangian.

The simplest such Lagrangian would contain only the SM degrees of freedom, together with additional operators beginning at dimension
six. Such a Lagrangian would arise, for example, as the
low energy limit of the `minimal composite Higgs model' \cite{CH:ACP},
based on the coset $SO(5)/ SO(4)$. Clearly, since the SM on its own is deficient
from the point of view of EWBG, the dimension-six (or higher)
operators must have a large effect on both the EWPhT (via sextic and
higher contributions to the Higgs
potential~\cite{Grojean:2004xa,Delaunay:2007wb,Grinstein:2008qi}) and on
$CP$-violating physics~\cite{Bodeker:2004ws,Grinstein:2008qi}, but
this conflicts with the need for a large strong-coupling scale (which suppresses the higher-dimension operators)  and in
any case might jeopardize the validity of the effective field
theory expansion.

However, there does not seem to be any compelling reason ({\em pace Occam}) 
to choose
the composite Higgs model based on the coset $SO(5) /SO(4)$ over one
based on a larger coset, and featuring the same desirable
properties \cite{Gripaios:2009pe,Mrazek:2011iu}. On the contrary,
from the point of view of the EWPhT, we know that the most favorable
case (at the renormalizable level) occurs when the Higgs sector is
extended to include a gauge singlet scalar and that in this way, one
may have a strongly first-order phase transition (for a recent
comprehensive analysis, see \cite{USsinglet} and references therein).
Such a scenario is realized in the
composite Higgs model based on the global symmetry breaking pattern
$SO(6)\rightarrow SO(5)$ \cite{Gripaios:2009pe}, where the coset is
five-dimensional and the low-energy spectrum includes four degrees of
freedom corresponding to the Higgs doublet plus one, real, singlet
PNGB. What is more, the non-renormalizable operators in the low-energy
effective Lagrangian of this model begin at dimension-five and include
an operator coupling the singlet and the Higgs to a pair of top
quarks that violates $CP$.

In this article, we show that such a model can generate the
 baryon asymmetry. As we have argued, it suffices to study the
low-energy effective theory of the SM plus a singlet, including the
aforementioned dimension-five operator. The scenario offers a testable way to explain the origin of the baryon asymmetry and can also be compared with constraints on new, $CP$-violating physics coming from
electric dipole moment (EDM) tests and from LEP.

In Section~\ref{Sec:SingletModlel} we summarize the features of the composite Higgs model with a singlet that are relevant for baryogenesis (more details are given in Appendix~\ref{Sec:CH}) and in Section~\ref{sec:EWBG} we study how the
baryon asymmetry arises in this scenario. In Section~\ref{sec:EDM}, we
study electric dipole moment and LEP constraints, while
in Section~\ref{sec:SCPV} we quantify how much explicit $CP$ violation
is needed to obtain a sufficient net baryon asymmetry. In
Section~\ref{sec:PT} we estimate the characteristics of the phase
transition (such as the wall thickness and critical temperature) in a special case where the theory is approximately $\Z2$-symmetric: then the structure of the effective Lagrangian is simpler and allows for an analytical study. Finally, in Section~\ref{Sec:Concl} we present our conclusions. In Appendix~\ref{Sec:appendixTransport}, we collect the transport equations used to calculate the baryon asymmetry.


\section{The SM plus a Singlet from a Composite Higgs\label{Sec:SingletModlel}}

We are interested in composite Higgs models that, in the low energy spectrum, include the SM and a further real, scalar degree of freedom, singlet under the SM gauge group. One example is the model based on the $SO(6)/SO(5)$ coset of ref.~\cite{Gripaios:2009pe}, which we summarize in Appendix~\ref{Sec:CH}. In this Section, we highlight the features that play a r\^ole in EWBG: in particular, we concentrate on the effective scalar potential and on the couplings between the Higgs and top-quark sectors, which, from naturalness arguments, are expected to be mostly composite.

The most general effective scalar potential at the renormalizable level is
\begin{equation}\label{Vtree}
V= V^{\textrm{even}}+V^{\textrm{odd}}\ ,
\end{equation}
with
\bea
\label{VtreeEven}
V^{\textrm{even}}&\equiv& -\mu_h^2 |H|^2+\lambda_h |H|^4\nonumber \\ 
&& \, - \, \frac{1}{2}\mu_s^2 s^2 +\frac{1}{4}\lambda_s s^4
+\frac{1}{2}\lambda_m s^2|H|^2,\\ 
V^{\textrm{odd}}&\equiv &\frac{1}{2}\mu_m s |H|^2 +\mu_1^3 s
+\frac{1}{3}\mu_3 s^3\ ,\label{VtreeOdd}
\eea
where $\mu_{h,s,m,1,3}$ have dimension of mass and
$\lambda_{h,s,m}$ are dimensionless\footnote{The singlet extension of the SM can  produce a strongly first-order phase transition already at the
renormalizable level. So, provided $v$ and $\Delta s$ (the jump in $s$ at the EWPhT) are small compared to $f$, we can ignore higher dimension operators in the potential or the scalar kinetic terms.}; $H$ denotes the Higgs $SU(2)_L$ doublet with physical component $h/\sqrt{2}$. $V^{\textrm{even(odd)}}$ denotes
the part of the potential that is even (odd) with respect to the
$\mathbf{Z}_2$ transformation
\begin{equation}
s\rightarrow -s\ .
\end{equation}
While this is an isometry of the coset space, whether or not it is a
symmetry of the effective Lagrangian depends on how the SM fermions
are coupled to the sigma model~\cite{Gripaios:2009pe}.

Let us now consider the couplings between the singlet and the fermions.
Lorentz invariance alone allows the singlet $s$ to couple to a Dirac fermion $F$ via
\be
s \bar{F}(a+ib\gamma_5)F\ ,
\ee
where $a$ $(b)$ is a dimensionless
coefficient describing its (pseudo)scalar-like couplings. In the SM, however, the $SU(2)_L\times U(1)_Y$ gauge
symmetry forbids such a term in the Lagrangian and $s$ can interact
with the SM fermions only at the non-renormalizable level, beginning
at dimension five with the operator
\be\label{topcoupling}
\frac{s}{f} H \bar{Q}_3(a+ib\gamma_5) t + {\mathrm h.c.}\ ,
\ee
where $f$ is the analogue of the pion decay constant and is related to
the mass $m_\rho$ (of order the confinement scale $\Lambda$) and
coupling $g_\rho$ of the strong sector resonances via $m_\rho=g_\rho
f$, where $g_{SM}\lesssim g_\rho \lesssim 4\pi$ and $g_{SM}$ is a
typical SM coupling \cite{SILH}. In eq.~(\ref{topcoupling}) we have
written only the
coupling between the singlet $s$ and the third generation $SU(2)_L$ doublet, $Q_3$, and
singlet, $t$.
Indeed, naturalness implies that the Higgs and top
sectors be mostly composite, so that the strong dynamics is expected
to influence mostly the interactions within and between these two sectors. Even
in this case, interactions with the lighter fermions will be present
in the mass eigenstate basis, but are expected to be of the order of
the corresponding (small) Yukawa couplings.  

Finally, it is useful for what follows to consider how one may implement $CP$ in this context: If
$V^{\textrm{odd}}$ vanishes, $a=0$ and $b\neq0$, the singlet
behaves as a pseudoscalar and $CP$ is conserved; similarly for $b=0$ and
$a\neq 0$ the singlet is scalar-like and $CP$ is also conserved in the
Lagrangian. Other non-trivial choices inevitably violate $CP$.


\section{Electroweak Baryogenesis\label{sec:EWBG}}

Two conditions need to be fulfilled during the EWPhT in order to
create enough baryon/antibaryon asymmetry
\cite{Sakharov:1967dj}. First of all, $CP$ violation must be present
within the wall separating the broken from the unbroken phase. This
sources an excess of left-handed versus right-handed
fermions\footnote{With left-handed (right-handed) we mean
$q_L+\bar{q}_R$ ($\bar{q}_L+q_R$), where the subscript $L$ denotes the
$SU(2)_L$ doublet and $R$ the singlet.} in front of the wall which
is converted into a baryon versus antibaryon excess by
non-perturbative electroweak (sphaleron) processes. For this excess to
be conserved, these sphaleron processes must be quickly suppressed
within the broken phase. This brings us to the second condition: that
the EWPhT be strongly first-order (if $v_c \equiv \left<
h\right>|_{T_c}$ is the value of the Higgs VEV in the broken phase at
the critical temperature $T_c$, then this condition reads
$v_c/T_c\gtrsim 1$ \cite{sphaleron_washout}). Neither of
these conditions is fulfilled in the SM, as the $CP$ violation encoded
in the CKM matrix is too small and, anyway, the phase transition is
really a crossover \cite{SMCross}, given the lower bound on the Higgs mass from LEP.

The strength of the EWPhT in the SM plus a singlet has been thoroughly
studied \cite{USsinglet, Anderson:1991zb, Espinosa:1993bs, Huber:2000mg,
Profumo:2007wc, Ashoorioon:2009nf, Cline:2009sn}. Many analyses
concentrated on loop effects
involving the singlet, which enhance the cubic term $ETh^3$ in the
Higgs potential at finite temperature, while reducing the quartic
$\lambda_h h^4$ (at a given Higgs mass) that enters the above
condition $1\lesssim v_c/T_c\approx E/\lambda_h$. LEP bounds on the
Higgs mass, however, suggest that one singlet scalar is not enough, if it
contributes only via loop effects \cite{Espinosa:2008kw}. Furthermore,
it was recently pointed out \cite{DeSimone:2011ek} that magnetic
fields generated during the EWPhT might increase the sphaleron rate within
the broken phase, calling for even stronger phase transitions in
order to have successful baryogenesis. The strongest phase transitions
are achieved when the singlet contributes through tree-level effects,
i.e. when the tree-level potential for $H$ and $s$ is such that a
barrier separates the EW broken and unbroken phases (not necessarily
with vanishing VEV $\left< s \right> $ along the singlet direction) \cite{USsinglet}. Indeed, in the
case of a barrier generated only at loop-level, the jump in the Higgs
VEV is proportional to the critical temperature $T_c$ (times a loop
factor), and is hence constrained to be small at small temperature. In
the case of a tree-level barrier, on the other hand, the Higgs VEV at
the critical temperature depends on a combination of dimensionful
parameters in the potential and its effect can be present even at small
$T_c$ (and is enhanced by a small $T_c$ appearing in the denominator of
$v_c/T_c$). In what follows we will concentrate on this possibility,
assuming that the transition is strongly first-order and relying on
the analysis of \cite{USsinglet}, which studies strong phase
transitions induced by tree-level effects in the SM plus a singlet. One
important implication of scenarios with a tree-level barrier is that a strong
transition is necessarily accompanied by a variation of the singlet
VEV during the EWPhT. This can be understood by noticing that, were the singlet VEV constant, the potential would have
the same shape as the SM potential at tree-level and would have, therefore, no tree-level barrier.

When the EWPhT is strongly first-order, bubbles of the broken phase
nucleate within a universe in the unbroken phase and expand.
$CP$-violating interactions can then source EWBG within the wall
separating the two phases. In the composite version of the SM plus a singlet outlined in the previous section, with
non-vanishing, pseudoscalar couplings between singlet and fermions
[$b\neq 0$ in eq.~(\ref{topcoupling})], the source is provided by a
variation in the VEV of $s$. Indeed, from eq.~(\ref{topcoupling}), we
can write the top quark mass, which receives contributions from both
$h$ and $s$, as
\be
\label{topmass}
m_t= \, \frac{1}{\sqrt{2}}v \,\left[y_t   + (a+ib) \frac{w}{f}\right] \equiv |m_t| \, e^{i\Theta_t}, 
\ee
where  $y_t$ is the top Yukawa and we defined the VEVs
\be
v \equiv \left< h \right>, \quad w\equiv \left< s \right>\ ,
\ee
with $v=246$ GeV. At vanishing temperature, the phase $\Theta_t$ can be absorbed in a
redefinition of the top quark field and is thus unphysical; the only
effect of a non-zero $w$ is a shift between the top-mass and the
Yukawa coupling compared to the relation that holds in the
SM. However, at finite temperature and, in particular, during a phase
transition, $w$ may change: then the space-time-dependent complex
phase in the quark mass cannot be rotated away by a simple field
redefinition, since it would reappear in the kinetic term. For this $CP$
violation to source EWBG, $\Theta_t$, and therefore $w$, must
change during the EWPhT \cite{Joyce:1994zn,
Cline:2000nw}. (As noted above, while this change is
not guaranteed when the barrier is induced by loop effects, a change
in $w$ is always present when the barrier appears at
tree-level). Then, assuming that we know how $w$ and $v$ change along
the direction $z$ perpendicular to the bubble wall, we may write
\begin{equation}
m_t(z)=|m_t(z)| \, e^{i\Theta_t(z)},
\end{equation}
similarly to what has  been done previously for the two-Higgs doublet
model (2HDM) \cite{Fromme:2006cm,Cline:1995dg}. This space-dependence
results in different dispersion relations for particles and
anti-particles (in particular top quarks) in the bubble wall, which,
in the WKB formalism \cite{Joyce:1994fu,Cline:2000nw,Fromme:2006cm},
imply force terms in the transport equations. This, in turn, causes an
excess of left-handed anti-quarks in front of the wall, in the symmetric
phase, which biases sphaleron transitions and generates a net baryon
asymmetry.

\begin{figure}[ht]
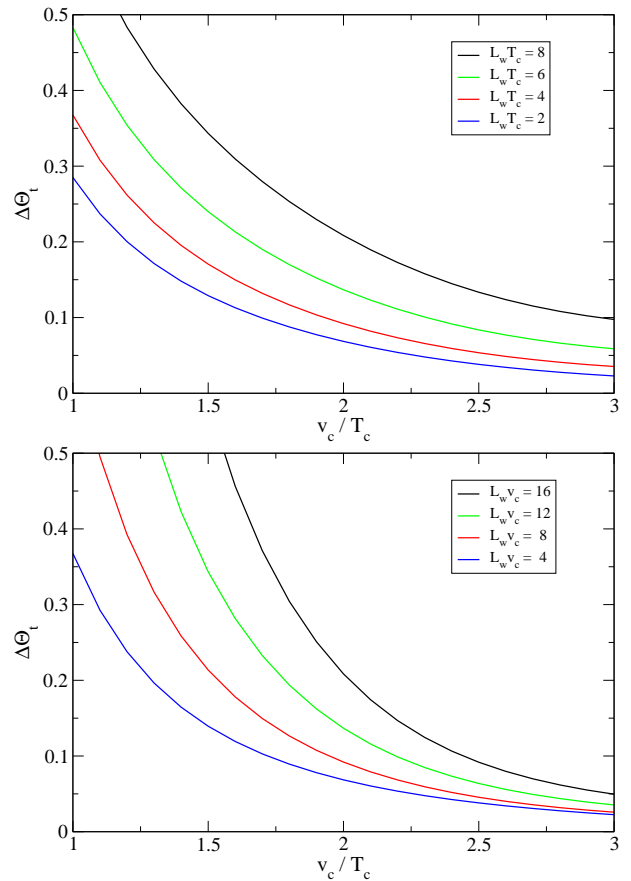

\begin{center}
\includegraphics[width=0.45\textwidth,clip]{figs/baryo_LT.eps}\\
\includegraphics[width=0.45\textwidth,clip]{figs/baryo_Lv.eps}\\
\caption{ \small The change in $\Theta_t$ needed during
the EWPhT to reproduce the observed baryon asymmetry
$(n_B-n_{\bar{B}})/n_\gamma\approx 6\times 10^{-10}$, as a function of
the strength of the phase transition $v_c/T_c$.  In the
top plot the wall thickness $L_w$ is fixed in units of the critical
temperature $T_c$ while in the bottom plot it is fixed in units of the
critical VEV $v_c$.}\label{DeltaS}
\end{center}
\end{figure}

We 
estimate 
the resulting asymmetry as follows. First, we
approximate both scalar field solutions to have a kink profile of
common extent along the $z$ direction, so that
\be\label{kinkansatz}
w(z) \equiv w_c + \frac{\Delta w_c}{2}\left[ 1+\tanh\left(z/L_w\right)\right],
\ee
and
\be
v(z) \equiv \frac{v_c}{2}\left[ 1+\tanh\left(z/L_w\right)\right],
\ee
where $L_w$ is the width of the wall, $\Delta w_c$ is the total change
in $\langle s\rangle$ at the critical temperature, $T_c$, and $v_c$ is the
Higgs VEV at $T_c$. The 
complex phase of the top mass then changes as
\be 
\Theta_t(z) \simeq \frac{\Delta \Theta_t}{2}
\left[1+\tanh\left(z/L_w\right)\right],
\ee
with
\be
\Delta \Theta_t \simeq \frac{b}{y_t} \frac{\Delta w_c}{f}. 
\ee
With this ansatz, we solve the transport equations numerically, using
the method of \cite{Bodeker:2004ws}, which we summarize for
completeness in Appendix~\ref{Sec:appendixTransport}. The results are displayed in
Figure~\ref{DeltaS}, where we show, for a given strength of the phase
transition and width of the wall (both are approximated analytically, for the $\Z2$ symmetric case, in Section~\ref{sec:PT}), the source of $\Delta \Theta_t$ that
is needed to reproduce the observed baryon asymmetry,
$(n_B-n_{\bar{B}})/n_\gamma\approx 6\times 10^{-10}$
\cite{Komatsu:2010fb}. The final baryon asymmetry increases with $\Delta \Theta_t$ or
$v_c/T_c$ but decreases with $L_w T_c$ or $L_w v_c$.

\begin{figure}[ht]
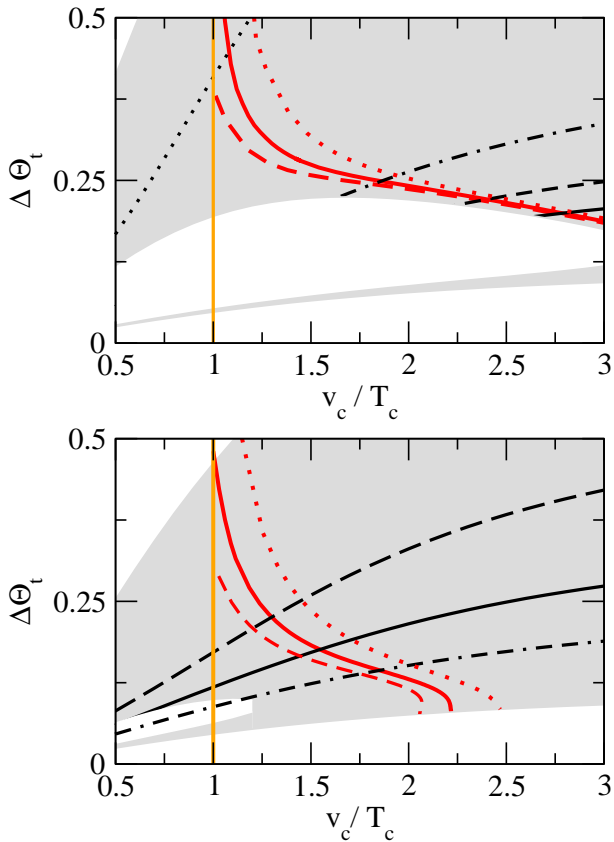

\begin{center}
\includegraphics[width=0.45\textwidth,clip]{figs/MS80.eps}\\
\includegraphics[width=0.45\textwidth,clip]{figs/MS130.eps}\\
\caption{ \small Shaded region: for $f/b=500\GeV$, $m_h=120\GeV$ and $m_s=80,\ 130\GeV$ (upper and lower plots), the $\Delta\Theta_t$ achieved for a given $v_c/T_c$ in the $Z_2$-symmetric case (a tiny explicit breaking is assumed, see Section~\ref{sec:SCPV}). The black lines (dotted, dot-dashed, dashed, solid, double dashed-dotted) correspond to explicit examples with fixed $\lambda_m=0.25,0.5,0.75,1,1.5$, respectively. Points on the red lines match the observed baryon asymmetry (solid) or 1.5 (dotted), 0.75 (dashed) times that value. The vertical line marks $v_c/T_c=1$, below which the asymmetry would be erased by active sphalerons.}
\label{DeltaSZ2}
\end{center}
\end{figure}

In Figure~\ref{DeltaSZ2} we show the change in $\Theta_t$ taking place during a strong phase transition in the version of this model with a $\Z2$ symmetry, as described in Sections~\ref{sec:SCPV} and \ref{sec:PT}. For $f/b=500\GeV$, $m_h=120\GeV$ and $m_s=80,\ 130\GeV$ (upper and lower plots respectively), the shaded region shows a projection of the other parameters of the model onto the $v_c/T_c$ vs. $\Delta\Theta_t$ plane. The lower and upper bounds on the shaded regions are due to the requirement of having the right EW minimum, eq.~(\ref{eq:cc_bound}), while the unshaded region in the middle correspond to an instability of the vacuum at the critical temperature, eq.~(\ref{kappak}). The black lines are examples with fixed $\lambda_m$, and along the red lines a fixed amount of baryon asymmetry is reproduced (the solid line corresponding to the observed amount).

To summarize, for a strong (tree-level) phase transition, as may occur
in the SM plus a singlet \cite{USsinglet}, a non-vanishing
pseudoscalar coupling $i\,b\gamma_5$ between $s$ and the top quark
generates a baryon asymmetry. For a strong enough EWPhT with $b\Delta
w_c / f\gtrsim 0.1$, the observed value of $(n_B-n_{\bar{B}})/n_\gamma$
can be reproduced. In the composite version of the SM plus a singlet, these requirements are fulfilled for natural values of the parameters.

We close this Section with a comparison of our EWBG scenario with previous studies of EWBG in non-supersymmetric models, such as the two-Higgs doublet model \cite{Fromme:2006cm,Cline:2011mm} or the SM with a low cut-off \cite{Grojean:2004xa,Bodeker:2004ws,Delaunay:2007wb,
Grinstein:2008qi}. In the former, $CP$ violation arises already at the
level of renormalizable operators in the Higgs potential, through a
complex phase between the two Higgs VEVs. Very strong phase
transitions (induced by tree-level barriers) are not possible in that
context since, contrary to the case with a singlet, the second Higgs
doublet cannot acquire a VEV prior to the EWPhT by definition. (To
circumvent this problem, ref.~\cite{McDonald:1993ey} studies a 2HDM
with an additional singlet: the two Higgs doublets violate $CP$; the
singlet strengthens the EWPhT.) Although the non-supersymmetric 2HDM
does not address the hierarchy problem, it is worth noting that it can also arise as the low-energy limit of composite Higgs models \cite{Mrazek:2011iu}.

The behaviour at finite temperature of other scenarios that address
the hierarchy problem but lead only to a light single Higgs, such as
the Minimal Composite Higgs \cite{CH:ACP} or Little Higgs models, have
been also analyzed. Refs.~\cite{Grinstein:2008qi} studied the
temperature behaviour of a Higgs that arises as the PNGB of a broken
global symmetry,\footnote{At even higher temperatures, the same
  mechanism that cuts off quadratic divergences in the Higgs potential
  also affects its finite temperature corrections and could lead to
  non-restoration of the EW symmetry \cite{Espinosa:2004pn}.}
parametrizing the deviations from the SM through effective
operators. A strong EWPhT can result in this setting from the
dimension-six operator $h^6$, which stabilizes a Higgs potential with
negative quartic coupling, as discussed in
\cite{Grojean:2004xa,Delaunay:2007wb}. This creates a large tree-level
barrier but the reliability of the effective-theory description is not
then obvious. Different dimension-six operators are responsible for sourcing $CP$ violation \cite{Bodeker:2004ws,Grinstein:2008qi}, in a manner similar to
our eq.~(\ref{topmass}), and for generating a complex mass for the
top quark: $m_t\sim y_t(v_h+iv_h^3/\Lambda^2)$. Compared to the model proposed here, these operators (which would arise also in our model, in the limit of a heavy singlet) are dimension-six and hence generally smaller than the ones involving the singlet.


\section{Electric Dipole Moments and other Constraints\label{sec:EDM}}

\begin{figure}[ht]
\begin{center}
\includegraphics[width=0.35\textwidth,clip=true]{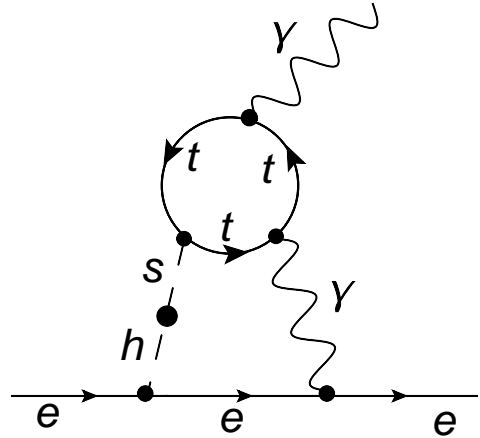}
\caption{\small Diagram illustrating the largest contribution to the
electron EDM: the dashed line indicates a Higgs that mixes with the
singlet, which then couples with the top.}
\label{BarrZeeE}
\end{center}
\end{figure}

The presence of a scalar that mixes with the Higgs and has
pseudoscalar couplings to fermions induces an electric dipole moment
(EDM) for the electron and for the neutron. 
The electron EDM receives the largest contribution from the
two-loop Feynman diagram \cite{Barr:1990vd} of Figure~\ref{BarrZeeE}, 
where the electron flips its chirality by coupling to the Higgs, which
then mixes with the singlet. The singlet, in turn, couples to two
photons through a top quark loop. We assume here that the
singlet couples only to the top quark, which is the only coupling
required for EWBG. If it couples also to light quarks, then other
diagrams may give larger effects \cite{Pospelov:2005pr}.\footnote{A similar diagram arises in
the 2HDM \cite{Fromme:2006cm}, except that there the singlet is
replaced by the $CP$-odd neutral scalar $A^0$. Diagrams with a $Z$
boson instead of a photon give suppressed contributions to EDMs
\cite{Barr:1990vd}.} Note that if
these couplings were hierarchical and if light flavors couple to the
singlet with a strength of order their Higgs Yukawa, as in the
scenario of partial compositeness \cite{PartComp}, then the diagram of
Figure~\ref{BarrZeeE} dominates. This diagram, which receives
opposite sign contributions from the two scalar mass eigenstates \cite{Barr:1990vd}, gives
\bea
\label{EEDM}
\frac{d_e}{e} &=&  Z\,\frac{16}{3}\frac{\alpha}{(4\pi)^3}\frac{m_e}{v^2} \nn \\
 && \times \,\frac{b}{y_t}\,\frac{v}{f}\Big[G\left(m_t^2/m_1^2\right)-G\left(m_t^2/m_2^2\right)\Big],
\eea
where $m_{e,t}$ are the electron and top quark masses, $Z$ is the mixing between $h$ and $s$ \cite{Weinberg:1990me}, given by
\begin{equation}\label{Z}
Z=\cos\theta\sin\theta=\frac{m_{sh}^2}{\sqrt{(m_h^2-m_s^2)^2+4m_{sh}^4}}\ ,
\end{equation}
and
\begin{equation}
G(z)\equiv z \int_0^1dx \frac{1}{x(1-x)-z}\log\left[\frac{x(1-x)}{z}\right]\ .
\end{equation}
The angle $\theta$ measures the mixing between the ($h,s$) gauge
eigenstates and the mass eigenstates ($\eta_1,\eta_2$), with masses
$m_{1,2}$:
\begin{eqnarray}
\eta_1&=s \sin\theta+h \cos\theta\ ,\\
\eta_2&=s \cos\theta-h \sin\theta\ .
\end{eqnarray}
The mostly-doublet eigenstate is $\eta_1$, so that $\theta\in
[-\pi/4,\pi/4]$. Finally, $m_s^2$, $m_h^2$ and $m^2_{sh}$ are the
entries of the (squared) mass matrix  for $s$ and $h$ at the electroweak minimum (their
expression in terms of the parameters of the potential
of eq.~(\ref{VtreeOdd}) can be found in \cite{USsinglet}).

The present bound on the electron EDM~\cite{Hudson:2011zz} (at 90\%
C.L.) is
\begin{equation}
\frac{d_e}{e}<1.05 \times 10^{-27}\textrm{cm}\ ,
\end{equation}
leading to the constraint
\begin{equation}\label{constraint}
Z \, \frac{b}{y_t}\, \frac{v}{f}\left|G\left(m_t^2/m_1^2\right)-G\left(m_t^2/m_2^2\right)\right|\lesssim 0.32\ ,
\end{equation}
which is not very restrictive, since $|Z|<1/2$ and $v\ll f$, while the absolute value of the difference is smaller than~$\sim 1$ for $m_s\gtrsim 40 \GeV$.

A similar diagram \cite{Gunion:1990iv}, with an up or a down quark
$q_{u,d}$ replacing the electron, and with a gluon replacing the
photon, induces a chromoelectric dipole moment (CEDM) $d_{u,d}^{C}$
for the up and down quarks, defined via
\begin{equation}
\mathcal{L}\supset -\frac{i}{2}d^{C}_{u,d}\,g_s\,\bar{q}_{u,d}\,\sigma_{\mu\nu}\tilde{G}^{\mu\nu}q_{u,d}\, .
\end{equation}
This diagram gives the largest contribution to the neutron EDM
\cite{Pospelov:2005pr} in our scenario,
\begin{equation}\label{errorCEDM}
\frac{d_n}{e}\approx(1.1\pm0.5)\,\frac{d^{C}_d+0.5d^{C}_u}{g_s(1\GeV)}.
\end{equation}
The computation of the contribution is similar to that which leads to
eq.~(\ref{EEDM}). In the end, one finds \cite{Gunion:1990iv},
\bea
\frac{d^{C}_{u,d}}{g_s(1\GeV)} &\approx& \frac{d^e}{e}\frac{9}{8}
\frac{\alpha_s(m_t^2)}{\alpha_{em}(m_t^2)}  \left[\frac{\alpha_s(m_t^2)}{\alpha_s(1\GeV^2)}\right]^{14/23}\frac{m_{u,d}}{m_e}\, .\nn
\eea
Plugging in the numbers and comparing with the present bound on the
neutron EDM (which, at 90\% C.L. \cite{Nakamura:2010zzi}, is
$d_n/e<2.9\times 10^{-26}\textrm{cm}$), one obtains a stronger constraint:
\begin{equation}\label{constraint_b}
Z \, \frac{b}{y_t}\, \frac{v}{f}\left|G\left(m_t^2/m_1^2\right)-G\left(m_t^2/m_2^2\right)\right|
\lesssim (0.06-0.19),
\end{equation}
where the uncertainty stems from uncertainties in the measured values of the up and down quark masses \cite{Nakamura:2010zzi} and from eq.~(\ref{errorCEDM}).

\begin{figure}[ht!]
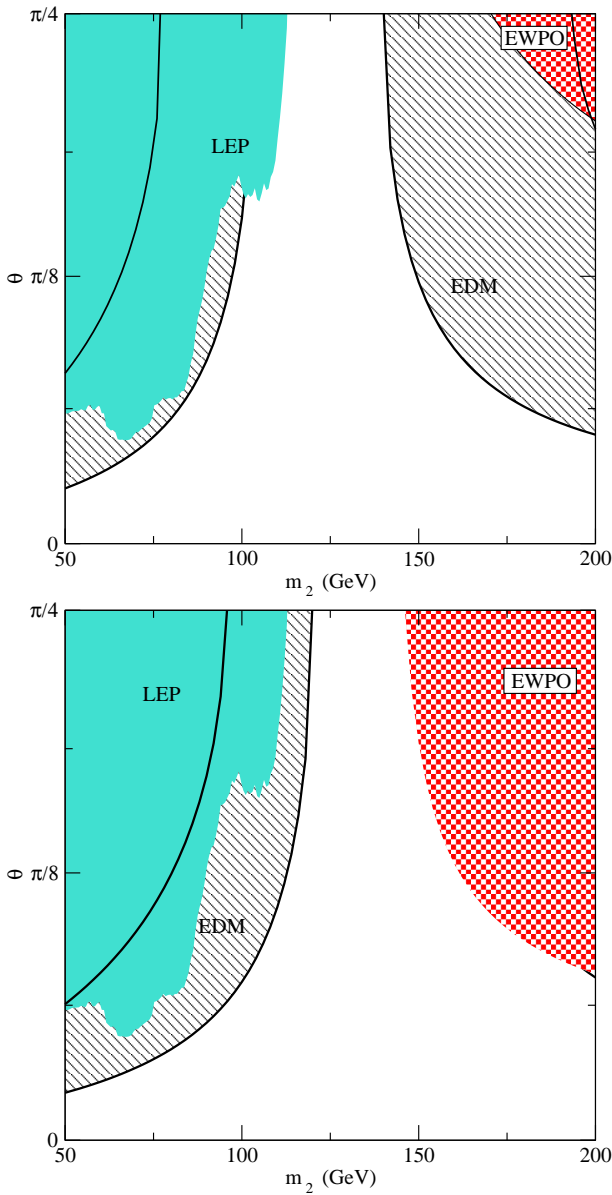

\begin{center}
\includegraphics[width=0.45\textwidth,clip]{figs/EDMmh120.eps}
\includegraphics[width=0.45\textwidth,clip]{figs/EDMmh140.eps}
\caption{\small Bounds on the mass $m_2$  of the mostly-singlet mass eigenstate, and the mixing angle 
$\theta$ (shown only in the range $0$ to $\pi/4$ since the limits do not depend on the sign of $\theta$). EWPO exclude, for $m_1=120,\ 140\GeV$ (upper and lower plot respectively), the red region. Direct searches at
LEP exclude the uniform turquoise  region for light $m_2$. EDMs, for $f/b=500\GeV$ exclude the striped region [the uncertainty (\ref{constraint_b}) in the boundary of this region is reflected
in the two limiting solid lines shown].}\label{fig:bounds}
\end{center}
\end{figure}

Strong constraints on the presence of a singlet scalar that mixes with
the Higgs also come from LEP, in the form of direct searches
\cite{Barate:2003sz} and electroweak precision observables (EWPO)
\cite{Ashoorioon:2009nf,Lebedev:2011aq}. The former apply only when
one of the mass eigenstates is lighter than $114 \, \GeV$: in this case,
direct searches constrain its mixing with the Higgs $h$, parametrized
by the angle $\theta$ in eq.~(\ref{Z}). In addition, EWPO constrain
the contribution of heavy states to the $\rho$ parameter, and also
lead to a bound on a combination of the mixing angle $\theta$ and the
mass eigenvalues. Neglecting the contributions of heavy composite states,  we can approximate
the latter bound, at 95\% C.L. \cite{GfitterNew}, as 
\begin{equation}
\cos^2\theta\ \ln m_{1}+\sin^2\theta\ \ln m_{2}<\ln 143 \GeV\ .
\end{equation}
We illustrate the various bounds in Figure~\ref{fig:bounds}, for a
representative scale of new physics given by $f/b=500\GeV$, as suggested
by the composite Higgs model of \cite{Gripaios:2009pe}, for Higgs
masses (here we mean the eigenstate that is mostly EW
doublet) $m_1=120,140\GeV$ (upper and lower plots). This choice of Higgs
masses avoids current LHC constraints, which might nevertheless become
relevant for other mass values or in the future, when more luminosity
is collected. EDM bounds disfavor large mixing angles and singlet masses much larger/smaller than $m_h$. In both cases, a light singlet with large mixing with the Higgs is disfavored by direct searches, as illustrated by the uniform turquoise region.


\section{Spontaneous $CP$ violation\label{sec:SCPV}}

In previous sections we have argued that in composite Higgs models
with a singlet coupled as a pseudoscalar to the top quark [eq.~(\ref{topcoupling})], the baryon asymmetry can be generated during the EWPhT. A necessary condition for EWBG is that the singlet VEV changes during the EWPhT, cf. Figure~\ref{DeltaS}. However, this condition is not sufficient, as we will show in this section. We consider a scenario
in which the scalar potential is $\mathbf{Z}_2$-symmetric
[$V^{\rm odd}=0$ in eq.~(\ref{Vtree})] and the singlet coupling to fermions is purely pseudoscalar-like [$a=0$ in
eq.~(\ref{topcoupling})], so that the singlet is a $CP$-odd
field. This corresponds to a scenario without explicit $CP$ violation, but where $CP$ can be spontaneously broken by the VEV of $s$. However, as we will discuss below, the existence of the symmetry $s\rightarrow-s$ leads to a cancellation between positive and negative contributions to the baryon asymmetry. In this section we study how a small explicit breaking of this symmetry can lead to a sufficient net baryon asymmetry. We note in passing that the case
$V^{\textrm{odd}}=0$ has been discussed in several contexts before: in a Singlet-Majoron model \cite{Cline:2009sn}, in the context of singlet dark matter \cite{Gonderinger:2009jp}, and in its own right~\cite{Profumo:2007wc}.

In the $\Z2$ symmetric case, a singlet VEV after the EWPhT produces dangerous cosmic domain walls \cite{Zeldovich} and is therefore excluded. Since, for baryogenesis, the singlet VEV must change during the EWPhT, $b\Delta w_c \neq0$, we are led to the case in which the singlet has a VEV prior to the EWPhT and none afterwards; this is fortunate since it is precisely the setting that produces the strongest phase transitions \cite{USsinglet}. In this case, there are in fact two transitions: first, the $\Z2$-breaking transition, at $T_s$, from the
point $(h,s)=(0,0)$ to some point $(h,s)=(0,\pm w_c)$  and 
later on, at $T_c<T_s$, the EWPhT from $(h,s)=(0,w_c)$ to $(h,s)=(v_c,0)$. Note that this sequence is quite plausible, as the Higgs, unlike $s$,
couples sizably to the gauge bosons and strongly to the top quark and therefore its mass receives larger finite-temperature corrections, which can delay its phase transition. 

Since the discrete $\Z2$-symmetry is broken spontaneously, albeit temporarily, at $T_s$, domain
walls (DWs) will be produced: if they live long enough, they might come to dominate the energy density of the universe, leading to a period of
late inflation and features in the cosmic microwave background which are in conflict with observations \cite{Zeldovich}. The temperature at which electroweak DWs start to dominate, however, is $T_{DW}\sim
10^{-7}\GeV$, which, in an expanding universe, happens much later than
the electroweak phase transition $T_{EW}\sim 100 \GeV$, when
the DWs disappear because the symmetry along the $s$ direction is
restored.

Issues with the vanishing of the total baryon asymmetry arise because,
during the first transition at $T_s$, different patches of the
universe (distributed equally, on average)  would end up in either the $+w_c$ or the $-w_c$ vacuum
(denoted $+$ and $-$ in the following), forming a network of domains. While the $+$ patches would
produce a net baryon asymmetry, the $-$ patches would produce a net antibaryon asymmetry (since the change in the singlet VEV during the EWPhT in different patches is equal and opposite in sign). In the following we discuss the possibility that a small explicit
breaking of the $\mathbf{Z}_2$-symmetry in the scalar potential
($V^{\textrm{odd}}\neq0$) lifts the degeneracy between the $\pm$
vacua and biases one vacuum with respect to the other so that, at the time of the EWPhT, the whole universe lies in the $-$ vacuum~\cite{McDonald:1995hp,
Sikivie:1982qv,DWdis,Abel}. We parametrize this
explicit breaking by a small difference $V(+w_c)-V(-w_c)=\Delta V$,
which can be generated, for instance, through loop effects involving a small $a\neq0$ in eq.~(\ref{topcoupling}). 

The network dynamics depends mainly on three forces: the pressure difference between the $+$ and $-$ phases $\Delta V$, the surface tension and the Hubble expansion. Friction is not very important, as discussed below. The surface tension $\sigma$ and the pressure difference balance each
other when the typical size of the domains $\xi$ is of the order of the size of the critical bubble for the transition $+\to-$, which scales as
\be
\label{eq:xi_crit}
\xi_{\rm cr} \sim \sigma/\Delta V\ .
\ee
As long as $\xi<\xi_{\rm cr}$ the dynamics of
the domains is dominated by the surface tension: small-curvature structures will be smoothed-out or collapse, increasing the typical length scale $\xi$ of the network and entering the so-called Kibble-regime, with $\xi$ growing linearly with conformal time. When $\xi$ reaches $\xi_{\rm cr}$, the pressure difference $\Delta V$ becomes
relevant and the energetically preferred $-$ regions expand exponentially fast into the $+$ regions. This behaviour has been confirmed by numerical simulations in different settings \cite{numDW}. In our model the temperature of the
electroweak phase transition is significantly lower than the
temperature of the first $\mathbf{Z}_2$-breaking phase transition (see next
section) so that DWs will evolve for a period of time
of order of the Hubble scale and all $+$ regions will have vanished at the time of the electroweak phase transition as long as $\xi_{\rm cr}\ll 1/H$.  This leads to a very weak lower bound on $\Delta V$ (for more details see~\cite{McDonald:1995hp})
\be
\label{boundDV}
\frac{\Delta V}{T^4}
\gg \frac{ H}{T}
\sim 10^{-16}.
\ee

Notice that this bound is in principle independent of the nature of the $\mathbf{Z}_2$-breaking phase transition. Indeed, in a first-order 
transition the typical size of domains is given
by the duration of the phase transition, which is of
order $\xi \sim 1/(100H)$~\cite{Anderson:1991zb}.  On the other
hand, if the transition is second-order, the typical size of
the domains is of order $1/T_G$~\cite{Kibble}, where $T_G\simlt T_c$
is the Ginzburg temperature  below which the broken phase settles in the either the $+$ or $-$ vacuum and thermal transitions to the other vacuum are Boltzmann suppressed \footnote{It has been suggested that the DW scale  is
  instead governed by a non-equilibrium mechanism related to critical
  slowing down~\cite{Zurek} leading to $\xi_Z\sim T_c^{-1}
  (M_{pl}/T_c)^{1/4}$ (which, for $T_c\ll M_{Pl}$, implies $1/T_G\ll
  \xi_Z\ll 1/H$). However, this analysis seems to require
  $(1-T_G/T_c)/(H\xi_0)^2<{\cal{O}}(1)$~\cite{Rivers} (where $\xi_0$
  is the field correlation length at $T=0$), a condition that is grossly violated for $T_c\ll M_{Pl}$.}. So, even if in most cases
one expects the pressure difference to dominate over the surface tension right after the end of the $\mathbf{Z}_2$-breaking phase transition,
if condition (\ref{boundDV}) is satisfied the DW length scale will reach $\xi_{cr}$, and DWs will quickly disapear,  before the EWPhT  \footnote{If  $T_G$ is high enough, it is possible to avoid the formation of DWs altogether. Indeed if the breaking is so big that, by the time the  + minimum develops, at $T_+$, thermal fluctuations to it are suppressed  ($T_+<T_G$). Assuming that the dominant $\Z2$-breaking comes from a tadpole, $\delta V=\mu_1^3 s$, the condition for $T_+<T_G$ to happen is:
\be
\mu_1^3>\frac{2}{3\sqrt{\lambda_s}}\frac{\mu_s^3}{(1+\frac{c_s}{128\lambda_s^2})^{3/2}}\ ,
\ee
which is much larger than (\ref{boundDV}).}.

Notice that it is also possible that DWs be still around at the time of the
EWPhT but that the volume occupied by the $-$ phase, $\mathcal{V}_-$, is sufficiently larger than the volume $\mathcal{V}_+$ of the  $+$ phase, so that the suppression factor $(\mathcal{V}_--\mathcal{V}_+)/(\mathcal{V}_-+\mathcal{V}_+)$ on the net baryon asymmetry produced is not particularly small. Interestingly, this can happen in a time scale significantly shorter than the one needed for domain wall collapse \cite{DWsV}, leading to a bound on $\Delta V$ even weaker than (\ref{boundDV}).

Finally, let us examine why friction is not important in our setting~\cite{McDonald:1995hp}, so that one expects DWs to move with relativistic velocities. Friction is due to the plasma 
having different properties in the two phases. For example, during the EWPhT, W-bosons and top quarks get a mass from their couplings to the Higgs VEV. This change in mass implies a change in momentum when these particles cross the Higgs bubble wall and this produces a friction force on the wall. Now however, the
properties of the particles in the $+$ and $-$ phases differ only by very small effects from the operators that induce the potential difference $\Delta V$. This suppresses friction and both sides of the equation
\be
\label{eq:v_term}
\Delta V \simeq v_w \, F_{\rm friction},
\ee
that determines the terminal velocity of the wall, $v_w$, vanish in the limit $\Delta V \to 0$. Quantum mechanical reflection effects of particles with an $s$-dependent mass that will feel the presence of the wall (as discussed in a similar context in~\cite{Abel}) are expected to be negligible too.

To go beyond this rough estimate, consider the approach of ref.~\cite{Dine:1992vs}, which works in the limit of a weakly coupled plasma and small wall velocities. The friction due to the mass change of the particles in the plasma between the phases is~\cite{McDonald:1995hp}
\be
F_{\rm friction} \sim \frac{1}{4\pi^2} \frac{T}{m}
\left[ 1 + \frac18 \log \left( \frac{4 m^2}{\delta m^2} \right) \right]
(\delta m^2)^2.
\ee
Since the pressure difference as well as the mass difference are
linear in $\Delta V$, this small source of friction cannot lead to small terminal wall velocities according to (\ref{eq:v_term}).

Nevertheless, in some cases with small friction one might end up with terminal velocities that are subsonic, as discussed in the analysis of~\cite{Konstandin:2010dm}, which is valid for strongly interacting plasmas (when the free streaming length of the particles in the plasma is much shorter than the wall
thickness). The hydrodynamic obstruction to a larger terminal velocity is due to the release of latent heat in front of the moving wall that increases the temperature and reduces, as a result, the pressure difference experienced by the wall. In principle, this effect can lead to small wall velocities when the pressure difference along the wall is much smaller than the latent heat. However, in the present case, both arise from the difference in potential $\Delta V$ so that the hydrodynamic obstruction (if present) has to occur at wall velocities that are of order of the speed of light.

To summarize, even though sources of explicit CP violation are needed in our setup~\cite{Cohen:1997ac}, already a very weak explicit breaking of CP in the scalar sector gives rise to substantial CP violation for electroweak baryogenesis. Essentially, this is due to
the two-stage nature of the phase transition.

In the composite Higgs model  on which we focus (Ref.~\cite{Gripaios:2009pe}, for details see Appendix~\ref{Sec:CH}),  a $\Z2$-preserving potential with a small explicit breaking could arise, for example, if the couplings of the top-quarks preserve the $\Z2$ [i.e. $a=0$ in eq.~(\ref{topcoupling})], while lighter quarks also couple to the singlet, but with $\Z2$-violating couplings. In this case the leading terms in the potential, coming from loop effects involving the top quark Yukawa, will generate $V^{even}$, while a small $V^{odd}$ would also be generated, proportional to the (smaller) light quark Yukawas. Alternatively, it might be possible that the same mechanism that is responsible for the CKM phase could generate a (presumably small) CP-violating contribution to the scalar potential via loop effects.

\section{Electroweak Phase Transition\label{sec:PT}}

In this section we discuss the characteristics of the phase transition
that are relevant for electroweak baryogenesis, namely the sphaleron
washout parameter $v_c/T_c$, the wall thickness $L_w$, and the change in singlet VEV during the EWPhT, $\Delta w_c$. Motivated by its simplicity and by the discussion of the last section, we focus on a scalar
potential with $\Z2$ symmetry, eq.~(\ref{VtreeEven}). We assume the existence of a small
odd part in the potential that breaks $CP$ explicitly, but this
contribution will hardly influence the transition dynamics.

Following~\cite{USsinglet}, we parametrize the potential in the mean-field
approximation as
\bea
\label{eq:Vcrit}
V(h,s,T) &=& \frac{\lambda_h}4 \left[ h^2 - v_c^2 + \frac{v_c^2}{w_c^2} s^2 \right]^2
+ \frac{\kappa}4 \,  s^2 h^2 \nonumber\\
&& + \frac{1}{2}(T^2 - T_c^2) (c_h \, h^2 + c_s \, s^2 )\ ,
\eea
where
\begin{equation}\label{kappak}
\kappa\equiv \lambda_m-2\lambda_h\frac{v_c^2}{w_c^2}\ .
\end{equation}
By construction, this potential leads to a tree-level barrier and hence a strong EWPhT. Indeed, at the critical temperature $T=T_c$, the potential has   degenerate minima at $(h,s)=(v_c,0)$ and
$(h,s)=(0,\pm w_c)$, separated by a barrier parametrized by
$\kappa>0$. The two coefficients $c_h$ and $c_s$ are, in the mean-field
approximation
\bea 
c_h &=& \frac1{48} \left[ 9 g^2 + 3 g^{\prime2} + 12 h_t^2 \right. \nn \\
&& \left. + \, \lambda_h (24 + 4 v_c^2/w_c^2) + 2 \kappa \right]\ , \\
c_s &=& \frac1{12} \left[ \lambda_h (3 v_c^4/w_c^4 + 4 v_c^2/w_c^2) + 2 \kappa \right]\ . 
\eea
The condition for ensuring that the electroweak minimum is the global
one at temperatures below $T_c$ reads
\be
\label{eq:cc_bound}
\frac{c_h}{c_s} > \frac{w_c^2}{v_c^2}\ .
\ee
This relation implies bounds on
$w_c$. Moreover, our low-energy effective theory approach also
requires $w_c \ll f $,  so that this parameter is typically restricted to a fixed range around $v_c$.

Given the form of the potential (\ref{eq:Vcrit}), the critical temperature $T_c$ can be expressed  in terms of $v_0$, the Higgs VEV in the EW breaking minimum at $T=0$, as
\be
\label{eq:Tcrit}
T_c^2 = \frac{\lambda_h}{c_h} (v_0^2 - v_c^2)\ .
\ee
This can be compared with the temperature at which the $\mathbf{Z}_2$-symmetry breaks:
\bea
T_s^2 &=& T_c^2 +  \frac{\lambda_h}{c_s} \frac{v_c^4}{w_c^2} \nn \\
&=& T_c^2 \left[ 1 + \frac{c_h}{c_s} 
   \frac{v_c^4}{w_c^2 (v_0^2 - v_c^2)} \right] 
> T_c^2 \frac{v_0^2}{v_0^2 - v_c^2}\ , 
\eea
which we have taken to be higher than $T_c^2$, as argued in the previous section.

The Higgs mass is 
\be
\label{eq:mh}
m_h^2 = 2 \lambda_h v_0^2\ ,
\ee
while for the singlet mass one finds
\be
\label{eq:ms}
m_s^2 = \frac12 \kappa v_0^2 + 
\lambda_h (v_0^2 - v_c^2) \left[ \frac{v_c^2}{w_c^2} - \frac{c_s}{c_h}\right]\ .
\ee
Both contributions to the singlet mass are positive, due to eq.~(\ref{eq:cc_bound}).\footnote{Very light singlets are disfavored: this is a consequence of our parametrization which, by construction, leads to tree level barriers. Light singlets can, of course, exist, but are generally not linked with very strong phase transitions (unless a cancellation in the parameters of the potential induces a flat direction \cite{USsinglet,DarkS}).}

Next, we determine the wall thickness in the thin-wall
approximation. The tunneling path will extremize the Euclidean action 
\be
S_1 = \int_{-\infty}^{+\infty} d\tau \left[\frac12 (\partial_\tau h)^2 + \frac12 (\partial_\tau s)^2
+ V(h,s)\right]\ , 
\ee
with boundary conditions
\bea
&&h(-\infty)=0\ ,\ h(\infty)=v_c\ ,\ h'(\pm\infty)=0\ ,\nonumber\\
&&s(-\infty)=w_c\ ,\ s(\infty)=0\ ,\ s'(\pm\infty)=0\ ,
\eea
and will proceed over (or close to) the saddle point with the potential value  
\be
V_\times = \frac14 \frac{\kappa \lambda_h v_c^4 w_c^2}
{(4 \lambda_h v_c^2 + \kappa w_c^2)}\ .
\label{Vsad}
\ee
We can estimate the parametric dependence of $L_w$ by approximating
\be
S_1\simeq \frac{\alpha}{L_w}(v_c^2+w_c^2)+\beta L_w V_\times\ ,
\ee
[where $\alpha,\beta\sim {\cal{O}}(1)$] which leads to 
\be
L_w^2=\frac{\alpha}{\beta}\frac{(v_c^2+w_c^2)}{V_\times}\ .
\ee
To estimate the numerical constant in this formula assume for
simplicity that the tunneling of the fields proceeds
along the path
\be
h(\tau) = v_c \, \sin[\phi(\tau)]\ , \quad
s(\tau) = w_c \, \cos[\phi(\tau)]\ ,
\ee
that passes a barrier of height $V = \kappa v_c^2 w_c^2 /16$. 
Using the ansatz
\be
\phi(\tau) = \frac{\pi}{4} [1 + \tanh(\tau/L_w)]\ ,
\ee
we obtain integral expressions for $\alpha$ and $\beta$ that can be calculated to give $\alpha/\beta\simeq 2.7$. In this way we arrive at the estimate 
\be
\label{eq:Lw}
L^2_w \simeq 2.7 \times \frac{1}{\kappa}\frac{v_c^2 + w_c^2 }{v_c^2 w_c^2}\left(1+\frac{\kappa w_c^2}{4\lambda_h v_c^2}\right)
\ . 
\ee
This result would become unreliable  for $\kappa\ll 1$, as the
potential then develops a flat direction and loop effects (giving rise to cubic terms from the Daisy
re-summed contributions) become important and strengthen the phase
transition \cite{USsinglet}. We do not discuss this particular case here.
Another pathological limit, $\lambda_h\rightarrow 0$, is forbidden by the lower limit on the Higgs mass.

To summarize, a scalar potential, its critical temperature and the wall width, are uniquely determined (in the $\Z2$ symmetric case) by the values of the Higgs and singlet masses at vanishing temperature, $m_h$ and $m_s$, by the strength of the EWPhT, $v_c/T_c$, and by the change in the singlet VEV, $w_c$. So, it is always possible to find parameter choices that give the right baryon asymmetry, as long as bounds for the consistency of the potential (such as perturbativity of the couplings, stability of the high-$T$  minimum  and of the $T=0$ EW breaking one) are respected: this is illustrated by the broad shaded regions of Figure~\ref{DeltaSZ2}.

We give some explicit examples in Table~\ref{tab:examples}. In the first case, we consider a model with small Higgs-singlet mixing  and a relatively light Higgs ($m_h \sim 120$ GeV). Hence, the experimental bounds of Figure \ref{fig:bounds} are fulfilled independently of the singlet mass. In the second case, we consider a model where the Higgs is relatively heavy ($m_h \sim 140$ GeV).  Besides this, we assume a sizable Higgs-singlet mixing, together with singlet and Higgs masses of similar size. This can be achieved by adding only a relatively small operator $s|H|^2$ to the scalar potential (that will have a negligible impact on the characteristics of the phase transition).  In both cases viable EWBG is possible with a strong
coupling scale $f$ as high as several TeV. Larger values of $f$ can be compensated by increasing $v_c/T_c$, which enhances the strength of the phase transition and the
CP-violating sources in the Higgs wall. However, notice that, although in our
simplified treatment, $v_c/T_c$ has been taken as a free parameter, for
very large values of $v_c/T_c$ ($\gtrsim 4$) one expects that the phase
transition will never be completed \cite{USsinglet}.

\begin{table}
\begin{tabular}{|c|c|c|c|c|}
\hline
& $\lambda_h$ & $\kappa$ & $w_c/v_c$ & $b w_c/f$ \\
\hline
\hline
S1 & $0.12$ & $0.12$ & $1.0$ & $0.1$ \\
\hline
S2 & $0.16$ & $0.48$ & $1.0$ & $0.15$ \\
\hline
\end{tabular}
\vskip 0.1 cm
\begin{tabular}{|c|c|c|c|c|c|c|}
\hline
& $m_h$ & $m_s$ & $v_c$  & $f/b$ & $L_w v_c$ & $v_c/T_c$\\
\hline
\hline
S1 & $120$ GeV & $81$ GeV  & $188$ GeV & $1.88$ TeV & $7.1$ & $2.0$\\
\hline
S2 & $140$ GeV & $139.2$ GeV  & $177.8$ GeV  & $1.185$ TeV & $3.5$ & $1.5$\\
\hline
\end{tabular}
\caption{\label{tab:examples} Numerical examples for models with
viable EWBG. The
entries of the first table are used as input while the second table is
produced using eqs.~(\ref{eq:Tcrit}), (\ref{eq:mh}), (\ref{eq:Lw}) and
Figure~\ref{DeltaS}.}
\end{table}

\section{Conclusions\label{Sec:Concl}}

Composite Higgs models have become very plausible options for natural
electroweak symmetry breaking and it is tempting to explore whether such scenarios can also solve other problems that the Standard Model fails to address.
One important and long-standing problem is the generation of the
cosmological baryon asymmetry and here we have shown in some detail how
EW baryogenesis can be successfully achieved in such a model.

The required ingredients turn out to be simple, not least because
naturalness itself implies modifications to the Higgs and top sectors,
which are precisely the ones believed to be responsible for
EWBG. Firstly, the electroweak phase transition can easily be strongly
first-order, for realistic values of the Higgs mass, if the model
contains in addition a scalar singlet. Moreover, if that scalar  couples to the top quark it can lead to a non-trivial $CP$-violating phase along the bubbles of the electroweak transition,
creating the seed for sphalerons to generate a non-zero baryon
asymmetry. We have shown that a composite model, based on the coset
$SO(6)/SO(5)$ proposed previously~\cite{Gripaios:2009pe}, contains all
these ingredients. Its low-energy effective theory features an
additional pseudo-Goldstone field that is a singlet and has
dimension-five, pseudoscalar couplings to top quarks that can break
$CP$. Motivated by this prospect, we have calculated quantitatively
the expected amount of asymmetry, showing that baryogenesis is
successfully achieved in a broad region of the parameter space that
passes other experimental constraints. Interestingly, the new complex phases and the mixing between the Higgs and the singlet predict contributions to
the EDMs of neutron and electron, not far beyond the reach of current and future experiments.

\section*{Acknowledgements}
Three of us (JRE, TK and FR) are grateful to the organizers of the Workshop 'Electroweak Baryogenesis in the Era of the LHC', at the Weizmann Institute, for stimulating discussions which shaped this article. JRE thanks T.~Hiramatsu, M.~Kawasaki and K.~'I.~Saikawa for sharing unpublished results of numerical simulations, while FR thanks  M.~Pietroni and A.~Riotto for interesting discussions. This work was supported by the Fondazione Cariparo Excellence Grant LHCosmo; the ERC starting grant Cosmo@LHC (204072); the Spanish 
Ministry MICINN under contracts FPA2010-17747 and FPA2008-01430; the Spanish Consolider-Ingenio 2010 Programme CPAN (CSD2007-00042); and the Generalitat de Catalunya grant 2009SGR894.  

\appendix

\section{The $SO(6)/SO(5)$ Composite Higgs Model}
\label{Sec:CH}
In composite Higgs models \cite{CH}, similarly to QCD, the hierarchy
between the Plank and the $\TeV$ scale is due to the slow logarithmic
running of an asymptotically free interaction that becomes strong and
confines close to the electroweak scale. New (techni-) fermions are
charged under this strong interaction but also under a global
symmetry; here we are interested in the case $SO(6)\simeq SU(4)$. As
the strong interaction confines, the global symmetry is broken down to
a subgroup, which we take to be $SO(5)$. From Goldstone's theorem,
we expect five massless degrees of freedom [equivalent to the number of
broken global symmetries in $SO(6)/SO(5)$], which are analogues of the
pions in QCD. The embedding of the SM group $SU(2)_L$ in $SO(5)$ is
such that these five Nambu-Goldstone bosons (NGBs), may be identified as
the four degrees of freedom belonging to the Higgs doublet plus one
degree of freedom, corresponding to a real singlet $s$. From symmetry
arguments alone, we can extract the leading kinetic part of the
$\sigma$-model Lagrangian (in the unitary gauge)
as~\cite{Gripaios:2009pe}
\bea
\mathcal{L}_{kin}&=&\frac{1}{2} (\partial_\mu h)^2+
\frac{1}{2} (\partial_\mu s)^2 \nn \\
&& + \,\, \frac{1}{2f^2}\frac{(h\partial_\mu h+s\partial_\mu s)^2}{1-h^2/f^2-s^2/f^2},
\label{kin}\eea
where $f$ is the analogue of the pion decay constant $f_\pi$, $h$ is
the physical Higgs and the fields are defined such that
\begin{equation}
\vev{h}^2+\vev{s}^2\leq f^2\,.
\end{equation}
The Higgs identified above transforms as a ${\bf (2,2)}$ of
$SU(2)_L\times SU(2)_R\sim SO(4)\subset SO(5)$ and its contributions to
the $\rho$-parameter are protected by the $SO(4)$ symmetry, which is
broken to the custodial $SO(3)$ symmetry in the EW vacuum.  The
$SO(6)$ symmetry is broken explicitly, both by the SM fermions, which
do not come in a complete representation of $SO(6)$, and by the gauging
of $SU(2)_L\subset SO(6)$. Thus, loops involving SM fermions or gauge
bosons communicate the explicit breaking to the (pseudo) NGBs and
generate a potential for the Higgs and the singlet.

In unitary gauge, the leading couplings between the $W^{\pm},Z$ gauge
bosons and the physical Higgs $h$ are
\begin{equation}
\mathcal{L}_{WZh}=\frac{g^2}{4}h^2\left[W^{\mu+} W^-_\mu+\frac{1}{2\cos^2\theta_W}Z^\mu Z_\mu\right]\, .
\end{equation}
We will assume the partial compositeness scenario
\cite{PartComp} (which seems most favorable from the point of view of
flavor physics), in which fermions couple linearly to operators of the
strong sector and embed them into representations of the global
$SO(6)$. As shown in \cite{Gripaios:2009pe}, among the possible
low-dimensional representations, only the $\mathbf{6}$ of $SO(6)$ 
can result in a realistic low-energy spectrum. The resulting couplings
of the fermions to the NGBs in the $\sigma$-model Lagrangian are
\begin{align}
\label{fermionCH}
\mathcal{L}_{\rm f} =&\sum_{j,k} \left\{\Pi_1^{j}(q^2)\,\bar q^j_L  \pslash q^j_L\, \frac{\redh^2}{f^2}\, \right.\nonumber\\
&+\Pi_2^{j}(q^2)\, \bar q^j_R \pslash  q^j_R \frac{1}{f^2}\left|  r(h,s)+ i \epsilon^j s\right|^2 \, \nonumber  \\  
&+\left. Y^{jk}(q^2) \,\bar q^j_L q^k_R\, \frac{\redh}{f} \left[r(h,s)
+ i \epsilon^k s \right]\,\right\}+ h.c. \, ,
\end{align}
with
\be
r(h,s)\equiv \sqrt{f^2-h^2-s^2}\ ,
\ee
and where the sum is over fermion flavors $j,k$ and $\epsilon^j$ are
complex numbers measuring the superposition between different
embeddings of the $SU(2)_L$ singlet quarks into representations of
$SO(6)$. Indeed, while the first four components of a $\mathbf{6}$ of
$SO(6)$ transform as a doublet under $SU(2)_L$, both the fifth and the
sixth transform as a singlet and can both accommodate the
$SU(2)_L$-singlet quarks. In eq.~(\ref{fermionCH}), $\Pi(q^2)$ and
$Y(q^2)$ are $q^2$-dependent form factors and depend on the details of
the UV physics. Note that these form factors give rise to the SM
Yukawas and, in particular, to interactions where $s$ couples to
the fermions as a $CP$-odd scalar (for a more detailed discussion on $CP$
in composite Higgs models see \cite{Gripaios:2009pe, Mrazek:2011iu}).

Loops involving SM fermions and gauge bosons break the global symmetry
$SO(6)$ explicitly and generate a potential for $h$ and $s$ which can,
in principle, be calculated if the form factors where known, or
estimated using the arguments of \cite{SILH}. The top quark and the
$W$-bosons, which have the largest couplings to the strong sector,
give the largest contributions, inducing a potential of the form
\bea
\label{pot}
V(\redh,\redeta) &\simeq& \, \, \alpha \redh^2+\lambda \redh^4   
+ \frac1{f^2} \left|  r(h,s)
+ i \epsilon^t \, s\right|^2 \nonumber \\ 
&& \hskip -1 cm
\times\left[\beta + \gamma \redh^2
+ \frac{\delta}{f^2}  \left|  r(h,s)
+ i \epsilon^t s\right|^2
\right]\, .
\eea
Here $\alpha, \lambda, \beta, \gamma,$ and $\delta$ are parameters
that depend on integrals over form factors, and $\epsilon^t$ is the $\epsilon$ parameter for the top quark and will be complex in general. This potential can be expanded
for small $s/f$ and $h/f$ in order to obtain the low-energy effective
potential. Notice that in the limit $\epsilon=\pm 1$, the potential does not depend on the singlet, which is, therefore, a massless NGB.

In summary, we see that the composite Higgs model of ref.
\cite{Gripaios:2009pe} based on the $SO(6)/SO(5)$ coset structure
reduces, at low energy, to the SM plus a singlet. The setup predicts
the right couplings to fermions [cf. eq.~(\ref{fermionCH}) versus
eq.~(\ref{topcoupling})] and the right potential [cf. eq.~(\ref{pot})
versus eq.~(\ref{Vtree})-(\ref{VtreeOdd})] to provide all necessary
ingredients to produce the observed baryon asymmetry through EWBG.
Furthermore, small explicit $CP$ violation (whose importance we discuss
in Section~\ref{sec:SCPV}) can be naturally understood in this
framework if $\epsilon^t$ is purely real or imaginary. In this case, the
dominant contributions to the potential for $s$ and $h$ preserve $CP$
and generate a $\mathbf{Z}_2$-symmetric potential of the form of
$V^{\rm even}$ in eq.~(\ref{VtreeEven}). In this case, the potential would
receive a subdominant $\mathbf{Z}_2$-violating contribution, $V^{\rm
odd}$, from  lighter quarks with complex $\epsilon$.


\section{Transport Equations}
\label{Sec:appendixTransport}

In our analysis of the baryon asymmetry, we adopt a set of
diffusion equations as in ref.~\cite{Bodeker:2004ws}.
The equations describe the evolution of (the $CP$ odd parts of) the
chemical potentials and velocity fields, denoted $\mu_i$ and $v_i$.
Initially, the species under consideration are the left-handed quarks,
the right-handed top and the Higgs. The right-handed bottom is
irrelevant since it is only produced by a chiral flip of the
left-handed bottom suppressed by the bottom mass. The dynamics
of the Higgs typically has a relatively small impact on the final
asymmetry and can be neglected~\cite{Fromme:2006wx}.

The diffusion equations for the left-handed quarks then read
\begin{eqnarray}\label{eq1}
&& \hskip -1 cm
(3\kappa_t+3\kappa_b)v_{\rm w}\mu_{q_3}^\prime
-(3K_{1,t}+3K_{1,b})v_{q_3}^\prime \nn \\
&& \hskip -1 cm
-6\Gamma_y\left(\mu_{q_3}+\mu_{t}\right)
-6\Gamma_m\left(\mu_{q_3}+\mu_{t}\right) \nn \\
&& \hskip -1 cm
-6\Gamma_{ss}\left[\left(2+9\kappa_t+9\kappa_b\right) 
\mu_{q_3}+\left(1-9\kappa_t\right)\mu_{t}\right] \nn\\
&=& 0 \ ,  
\end{eqnarray}
and
\begin{eqnarray}
&& \hskip -1 cm
-(K_{1,t}+K_{1,b})\mu_{q_3}^\prime 
+(K_{2,t}+K_{2,b})v_{\rm w}v_{q_3}^\prime  \nn \\
&& \hskip -1 cm
-\left(\frac{K_{1,t}^2}{\kappa_t D_{Q}}+ \frac{K_{1,b}^2}{\kappa_bD_{Q}}\right)v_{q_3}\nn\\
&=& K_{4,t}v_{\rm w}m^2_t\Theta_t'' 
+K_{5,t}v_{\rm w}(m^2_t)'\Theta_t' \ , 
\end{eqnarray}
while the diffusion equations for the right-handed top are
\begin{eqnarray} 
&&  \hskip -1 cm
3\kappa_tv_{\rm w}\mu_{t}'-3K_{1,t}v_{t}' \nn \\
&&  \hskip -1 cm
-6\Gamma_y\left(\mu_{q_3}+\mu_{t}\right)
-6\Gamma_m\left(\mu_{q_3}+\mu_{t}\right)\nn\\
&&  \hskip -1 cm
3\Gamma_{ss}\left[\left(2+9\kappa_t+9\kappa_b\right)
\mu_{q_3}+\left(1-9\kappa_t\right)\mu_{t}\right]\nn\\
&=& 0 \ ,
\end{eqnarray}
and
\begin{eqnarray}
&&  \hskip -1 cm
-K_{1,t}\mu_{t}'+K_{2,t}v_{\rm w}v_{t}'-\frac{K_{1,t}^2}{ \kappa_t D_Q}v_{t} \nn\\ \label{eq4}
&=& K_{4,t}v_{\rm w}m^2_t\Theta_t'' +K_{5,t}v_{\rm w}(m^2_t)'\Theta_t' \ ,
\end{eqnarray}
where primes denote derivatives w.r.t. the coordinate $z$ perpendicular to the wall. 
Compared to the implementation in~\cite{Bodeker:2004ws} we neglected some sub-leading
source terms that contain derivatives acting on the chemical
potentials and velocity fields (more precisely their $CP$-even parts).

The final baryon asymmetry results from the sphalerons acting on the
$CP$-odd component of the left-handed quark density
\bea
\eta_B &=& \frac{n_B}{s} =  
\frac{405 \Gamma_{ws}}{4 \pi^2 \, g_* \, v_w \, T^4} 
\int_0^\infty dz \, \mu_L(z) \, e^{-\nu z}\ ,  \label{eq:ws_int}  \\
\mu_L &=& (1 + 2 \kappa_t + 2 \kappa_b) \mu_{q3} - 2 \kappa_t \mu_t\ ,  \\
\nu &=& \frac{45 \Gamma_{ws}}{4 v_w T^3}\ , \quad g_* = 106.75\ .
\eea

The functions $K_{m,j}$ and $\kappa_i$ denote certain moments in
momentum space defined by
\bea
\left< X \right> &\equiv& \frac{\int d^3p \, X }
{\int d^3p \, f_+'(m=0)}\ , \nn \\
f_\pm(m_i) &=& \frac1{e^{\beta \sqrt{p^2 + m_i^2}} \pm 1}\ ,
\eea
namely
\bea
\kappa_i &\equiv& \left< f_\pm'(m_i)\right>, \nn \\
K_{1,i} &\equiv& \left< \frac{p_z^2}{\sqrt{p^2 + m_i^2}} f_\pm'(m_i)\right> \, , \nn \\
K_{2,i} &\equiv& \left< \, p_z^2 \, f_\pm'(m_i)\right> \, ,\nn \\
K_{3,i} &\equiv& \left< \frac{1}{2\sqrt{p^2 + m_i^2}} f_\pm'(m_i)\right> \, , \nn \\
K_{4,i} &\equiv& \left< \frac{|p_z|}{2(p^2 + m_i^2)} f_\pm'(m_i)\right> \, , \nn \\
K_{5,i} &\equiv& \left< \frac{|p_z| p^2}{2(p^2 + m_i^2)} f_\pm'(m_i)\right> .
\eea
The momenta are normalized such that for massless bosons (fermions)
one finds $\kappa_i=2(1)$. 

For the interaction rates and the quark diffusion constant we use the
values \cite{Huet:1995sh, weak_sph, Moore:1997im}
\bea
\Gamma_{ws} = 1.0 \times 10^{-6} T^4\ , && \quad 
\Gamma_{ss} = 4.9 \times 10^{-4} T^4\ , \nn \\ 
\Gamma_{y} = 4.2 \times 10^{-3} T\ , && \quad 
\Gamma_{m} = \frac{m_t^2}{63 T}\ , \nn \\
 D_Q &=& \frac{6}{T}\ .
\eea

As long as the wall velocity is clearly subsonic ($v_w \ll
1/\sqrt{3}$), the $CP$-odd particle densities are linear in the wall
velocities. If at the same time the sphaleron process is not saturated
(i.e. as long as the exponent in the integrand of (\ref{eq:ws_int}) is
small) the final asymmetry will depend only weakly on the wall
velocity. We assume that we are in this window (which might well be
due to hydrodynamic obstructions \cite{Konstandin:2010dm}) and adopt
the value $v_w=0.01$.

In our setup, the functions $K_{m,i}$, $\kappa_i$ and $\Gamma_y$
inherit a spatial dependence from the top mass. This precludes a
straightforward numerical solution of the system of diffusion
equations. We avoid this problem by matching the eigenvectors of the
diffusion system with the correct sign [i.e. the ones that die off at
infinity] far away from the wall, where all coefficients are constant,
to numerical solutions in the wall.

Ultimately, the baryon asymmetry is linear in the change of the phase
of the top mass $\Delta \Theta_t$ and otherwise only depends on the
dimensionless combination $v_c/T_c$ (that enters
in the change of the top mass during the phase transition) and the
wall thickness $L_w T_c$. Our numerical results are summarized in
Figure~\ref{DeltaS}.


\end{document}